%% file: lrec-coling2024-example.tex
\PassOptionsToPackage{table}{xcolor}
\documentclass[10pt, a4paper]{article}

\usepackage{lrec-coling2024}
\usepackage{natbib}
\usepackage{multibib}
\def\@mb@citenamelist{cite,citep,citet,citealp,citealt,citepalias,citetalias}

\usepackage{graphicx}
\usepackage{tabularx}
\usepackage{soul}
\usepackage{amsmath}
\usepackage{color}
\usepackage{multirow}
\usepackage{multicol}

\usepackage{algorithm}
\usepackage{makecell}
\usepackage[table]{xcolor}
\definecolor{mygray}{gray}{0.8}
\definecolor{myblue}{RGB}{183,230,230}
\definecolor{mygreen}{RGB}{184,238,184}
\definecolor{myyellow}{RGB}{255,255,164}
\definecolor{mypurple}{RGB}{207,182,255}
\usepackage{tabularx}
\usepackage{hyperref}
\definecolor{darkblue}{rgb}{0, 0, 0.5}
\hypersetup{colorlinks=true, citecolor=darkblue, linkcolor=darkblue, urlcolor=darkblue}
\usepackage{xstring}
\usepackage{booktabs}

\title{VCEval: Rethinking What is a Good Educational Video and How to Automatically Evaluate It}

\name{\and \textbf{Xiaoxuan Zhu$^{1,*}$~\thanks{$^*$Equal contribution}, Zhouhong Gu$^{1,*}$, Sihang Jiang$^1$, Zhixu Li$^1$, Hongwei Feng$^1$, Yanghua Xiao$^1$}}

\address{$^1$Shanghai Key Laboratory of Data Science, School of Computer Science, Fudan University, China \\
         \{xxzhu22, zhgu22\}@m.fudan.edu.cn\\
         \{tedsihangjiang\}@gmail.com\\
         \{zhixuli, hwfeng, shawyh\}@fudan.edu.cn\\}

\abstract{
Online courses have significantly lowered the barrier to accessing education, yet the varying content quality of these videos poses challenges.
In this work, we focus on the task of automatically evaluating the quality of video course content.
We have constructed a dataset with a substantial collection of video courses and teaching materials.
We propose three evaluation principles and design a new evaluation framework, \textit{VCEval}, based on these principles.
The task is modeled as a multiple-choice question-answering task, with a language model serving as the evaluator.
Our method effectively distinguishes video courses of different content quality and produces a range of interpretable results.
}

\begin{document}

\maketitleabstract

\section{Introduction}

\input{sources/01.Introduction.06}

\section{Related Work}
\input{sources/02.RelatedWork.02}

\section{Preliminary}
\input{sources/03.ProblemFormulation.03}

\section{Video Course Evaluation Framework}
\label{sec:method}
\input{sources/04.VCEval.04}

\section{Dataset and Human Evaluation}
\input{sources/05.Dataset.02}

\section{Experiments}
\input{sources/06.Experiments.03}

\section{Analysis}
\input{sources/07.Analysis.02}

\newpage
\section{References}\label{sec:reference}
\bibliographystyle{lrec-coling2024-natbib}
\bibliography{lrec-coling2024-example}

\end{document}

%% file: sources/01.Introduction.06.tex
The advent of online courses has revolutionized the way of disseminating knowledge. 
According to the report of the Ministry of Education of the People's Republic of China~\footnote{http://www.moe.gov.cn/}, more than 950,000 teachers from 1,454 universities or colleges in China have been teaching 942,000 online courses as of 2020, and the number of online courses is still increasing rapidly. As per the available data from Statista\footnote{https://www.statista.com/}, the number of users of online learning platforms is expected to amount to 0.9 billion by 2027.
This has significantly lowered the barriers to knowledge acquisition for the masses, enabling them to readily access the knowledge imparted by content creators of the online courses.
Nonetheless, as a nascent form of user-generated content, the quality of online courses is highly inconsistent. 
High-quality video courses deliver knowledge and skills effectively, while low-quality ones may squander the user's time and lead to misconceptions.

\begin{figure}[t]
    \centering
    \resizebox{\columnwidth}{!}{
    \includegraphics{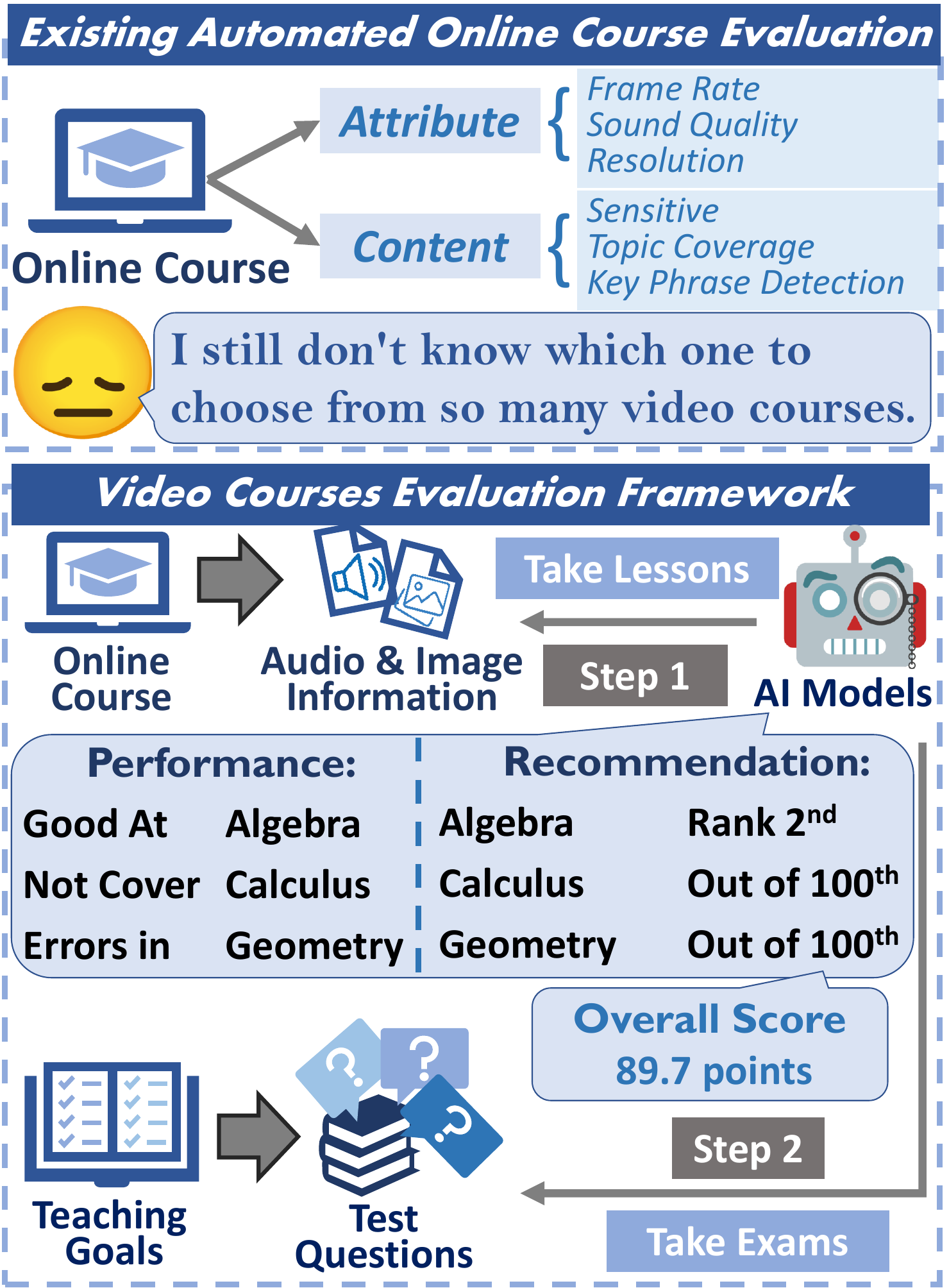}
    }
    \vspace{-3mm}
    \caption{\textit{Top:} Existing automated online course evaluation mainly focuses on the video attribute and video topic, failing to evaluate the video content's clarity in elucidating knowledge. \textit{Bottom:} Our proposed framework for automated evaluation of the teaching content in online courses.}
    \label{fig:intro}
    \vspace{-5mm}
\end{figure}

Nowadays, online learning platforms have incorporated automated evaluation methods for online courses to present certain evaluation metrics to users, content creators, and platform administrators~\citep{GiraldoGLSM23, DBLP:conf/aied/Zheng23, chen2022recommending}. 
However, these methods hardly focus on the content's clarity in elucidating knowledge. 
As shown in the top of Fig.~\ref{fig:intro}, the prevailing automated quality assessments predominantly fall into two categories:
(1) Video Attribute: This pertains to information concerning the video's frame rate, resolution, audio quality, and so forth~\citep{ahn2018deep, liu2018end, zhang2018blind, 9463703}.
(2) Video Topic: This pertains to the theme addressed in the video, the key content mentioned, the presence of sensitive information, and so forth~\citep{hasan-etal-2021-hitting}.

The provision of these metrics has been previously demonstrated to be effective in aiding users in selecting the requisite videos.
However, within the rapidly burgeoning realm of online education, the topic of the different video courses is becoming homogenized.
Based on user feedback from online learning platforms such as Coursera, MOOCs, and Bilibili, only providing these metrics, which are unrelated to the video content quality, no longer assists users in swiftly selecting courses that meet their needs from a vast array of videos.
Users often spend a large amount of time selecting and comparing similar courses.
Video uploaders want to find a way to improve their courses, while platform administrators wish to give more exposure to high-quality video courses to keep the User Retention Rate.
As a result, the evaluation of video course content quality has become crucial to meet all these needs.

Since video courses frequently encompass a substantial amount of textual information in both audio and visual form, text remains a prevalent method to impart knowledge.
A direct evaluation method would be to employ Large Language Models~(LLMs), such as ChatGPT~\citep{openaichatgpt} and GPT-4~\citep{openai2023gpt4}, which entails converting the video's auditory and visual content into text and using the large model to rate video courses directly based on the established evaluation criteria.
However, for all online learning platforms, utilizing existing large models for evaluation poses a considerable risk of data leakage when employing another company's API, unstable scoring due to the hallucination problem of generative models, inconsistency in evaluation owing to limited input length, and the substantial cost of constructing large models from the ground up.
Therefore, the research question arises: \textit{Can we devise a novel method to effectively and accurately evaluate the content quality of a video course?}

To address this challenge, a straightforward strategy, as shown in Fig.~\ref{fig:intro}, is to make models function as students, participating in these online courses and taking exams to output the exam score as the evaluation results.
However, this method poses three difficulties: 
(1). How to ensure the \underline{validity} of the generated scores?
(2). How to ensure the \underline{interpretability} of the evaluation results?
(3). How to ensure the \underline{fairness} of the evaluation process for different videos of the same category?

In this paper, we propose a \textbf{\textit{V}}ideo \textbf{\textit{C}}ourse \textbf{\textit{Eval}}uation~\textbf{\textit{(VCEval)}} framework for automated evaluation of the teaching content in online courses as shown in Fig.~\ref{fig:intro}.
Specifically, VCEval obtains a large number of associated teaching materials in line with the teaching targets using search engines or large models.
These teaching materials can be used to construct multiple-choice questions, ensuring the \underline{validity} of the test questions.
VCEval breaks down the teaching targets and constructs test questions for each sub-target so that the model's evaluation results are \underline{interpretable} by sub-target evaluation score.
Considering the instability of model learning and the disturbance brought about by random initialization, VCEval uses test questions to eliminate the model's prior knowledge, ensuring a \underline{fairer} evaluation for different videos of the same category.

Considering that there is currently no suitable benchmark to validate the effectiveness of video course evaluation methods, we propose a benchmark constructed using various video course targets at the first 12 years of a student's academic journey~(K12) in this paper. 
Since K12 video courses often have clear teaching objectives, which help to measure the teaching quality, we believe these types of videos are ideal for constructing benchmark data. Considering that video uploaders often upload multiple videos to form a series to explain certain topics, we have collected 15 series of open-source video courses. These 15 series cover three different subjects: geography, biology, and history, and include a total of 370 uploaded videos, with a total duration of 8753 minutes. We also collect the textbooks and syllabi related to these subjects and use Optical Character Recognition~(OCR) and Automatic Speech Recognition~(ASR) technologies to convert all the video contents into text. 

This paper makes the following contributions: 
\begin{itemize}
\item We propose a new standard for evaluating the quality of online courses, which is closer to the purpose of users watching teaching videos. 
\item We collect a large amount of data for evaluating the content quality of video courses, including videos, audio, keyframe images, transcribed text information, and multiple-choice test questions. 
\item We propose an innovative framework, VCEval, for automated evaluation of video course quality, which allows the language model to imitate students attending classes and then taking tests.
The experimental results demonstrate that VCEval exhibits an average consistency of 73.33\% with human annotators in terms of overall accuracy for the video course series.
It shows an average consistency of 76.87\% in the evaluation of specific teaching targets.
\end{itemize}

%% file: sources/02.RelatedWork.02.tex
\subsection{Video Content Understanding}
\input{sources/02.R_zyx}

\subsection{Teaching Materials Assessment}
We extensively referred to existing works on teaching materials assessment in the process of designing VCEval.
Teaching materials refer to those alternative channels of communication, which a teacher can use in the teaching and learning process to help achieve desired learning objectives \citep{amadioha2009importance}. \citet{zarqtouni2020use} develop an e-learning platform and use online formative assessment to improve material quality. \citet{doi:10.1080/10572252.2012.626690} propose an assessment rubric for evaluating online video tutorials based on a descriptive study of 46 teaching videos found on YouTube. \citet{FIORELLA2018465} focus on features of teaching videos (e.g. breaking lessons into segments paced by the learner) to study which feature can or can not improve learning. Although these works propose several evaluation metrics and methods, high human involvement is still required.
Instead, we propose an automatic method that is more efficient and lower-cost.

%% file: sources/02.R_zyx.tex
The first step in Video Course Evaluation involves comprehending the video content.
The most prevalent method for understanding video content is Video Captioning (VC), which employs a range of techniques to help better understand the video content~\citep{abdar2023review, moctezuma2022video, khurana2021video}.
Early works often rely on template-based approaches~\citep{kojima2002natural, guadarrama2013youtube2text, krishnamoorthy2013generating}, which lack flexibility and expressive modeling capacity. 
Since the emergence of deep learning, the encoder-decoder structure is widely utilized~\citep{pan2015hierarchical, venugopalan2015translating, wang2018reconstruction}, with a CNN often used in the encoder to extract visual features, and an RNN utilized in the decoding phase for language generation~\citep{olivastri2019end}.
With the success of transformer models~\citep{vaswani2023attention}, more recent works~\citep{ye2022hierarchical, gu2023text, Yamazaki0TRL23} start to use the transformer structure as part of the modules to improve performance.
Given that text is the most efficient and direct way to express knowledge, this paper also employs video captioning to understand the teaching content in online courses.

%% file: sources/03.ProblemFormulation.03.tex
\subsection{Problem Formulation}
\subsubsection{Video Course Quality Evaluation}
Video Course quality evaluation aims to assess the quality of hypothesis video courses $v$ in terms of certain target $t$~(e.g. knowledge about ancient England History).
\begin{equation}
    y = f(v,R(t)),
    \label{eq:vq_evaluation}
\end{equation}

where
(1) $v$ represents the video course to be evaluated.
(2) $t$ denotes the teaching or evaluation target.
(3) $R$ denotes to find the reference for the evaluation target.
(4) Function $f(\dot)$ could be instantiated as a human evaluation process or automated evaluation method.

\subsubsection{Meta Evaluation}

Meta-evaluation aims to evaluate the reliability of automated methods by calculating how well the automated scores ($y_{auto}$) correlate with human judgment ($y_{human}$) using correlation functions:
\begin{equation}
g(y_{auto}, y_{human}).
\end{equation}
In this work, we adopt two widely-used correlation measures:
(1) \textbf{Spearman} correlation $(\rho)$ measures the monotonic relationship between two variables based on their ranked values.
(2) \textbf{Pearson} correlation ($r$) measures the linear relationship based on the raw data values of two variables.

\subsubsection{Evaluation Methodology} 
The evaluation methodology delineates various aggregation techniques employed in the computation of correlation scores.
Specifically, consider a scenario where for each scored video clip $v_i, i \in [ 1, 2, \dots, n ] $ (for instance, video clips imparting diverse knowledge), there exist $J$ system outputs denoted as $v_{i,j}$, where $j\in[1,2,\dots,J]$.
The automatic scoring function is represented as $f_{auto}$ (for example, ROUGE), and the gold human scoring function is denoted as $f_{human}$.
For a specified evaluation target $t$, the meta-evaluation function $F$ can be formulated accordingly.

\textbf{Video-level Evaluation Methodology:} 
This methodology aims to illustrate the comprehensive discrepancy output by the function $f()$ on different videos $v_i$ with varying targets $t$ in comparison with human evaluation. 
\begin{equation}
\begin{aligned} 
F^{video}_{f_{auto},f_{human}}=\\
\frac{1}{n_1}\sum^{n_1}_{i=1}[g(\sum^{n_2}_{j=1}&f_{auto}(v_i,R(t_j)), \\
\sum^{n_2}_{j=1}&f_{human}(v_i,R(t_j)))]
\end{aligned}
\end{equation}

\textbf{Target-level Evaluation Methodology:} 
The target-level evaluation methodology stipulates that a correlation value is computed for each target independently based on the outputs of multiple systems, which is subsequently averaged across all targets.
\begin{equation}
\begin{aligned}  
F^{target}_{f_{auto},f_{human}}= \frac{1}{n}\sum^{n}_{i=1}\\
[g(f_{auto}(v,R(t_i)), &f_{human}(v,R(t_i)))]
\end{aligned}
\end{equation}
where $g$ can be instantiated as Spearman or Pearson correlation.

\subsection{Criteria for Educational Video Quality Evaluation}
In the context of online educational videos, we advocate for the following criteria to be employed in the assessment of their quality:

\textbf{Precision}
The evaluation methodology should yield a precise quantification of the video's quality.
This precision should encompass the accuracy of the information disseminated, the depth of the content explored, and the relevance of the content to the overarching topic.
This criterion can be represented as:
\begin{equation}
\underset{f_{auto}}{\arg\min}\ F^{video}_{f_{auto},f_{human}}
\vspace{-1mm}
\end{equation}

\textbf{Interpretability}
The evaluation methodology should be inherently interpretable.
The interpretation of the quality score should be made accessible to users, thereby aiding them in discerning the most valuable aspects of the video.
Furthermore, this interpretability should serve as a tool for video creators to enhance the quality of their content.
This criterion can be represented as follows: 
\begin{equation}
\underset{f_{auto}}{\arg\min}\ F^{target}_{f_{auto},f_{human}}
\vspace{-1mm}
\end{equation}

\textbf{Consistency} 
The evaluation methodology should be capable of generating consistent quality scores for videos.
This consistency should be maintained across different videos and over time, thereby ensuring a reliable assessment of video quality.
This criterion can be represented as follows:
\begin{equation}
\underset{f}{\arg\max}(f(V,R(t)) - f(V',R(t))), V' \subseteq V
\vspace{-4mm}
\end{equation}

%% file: sources/04.VCEval.04.tex
\begin{figure*}[ht]
\centering

\resizebox{\textwidth}{!}{
\includegraphics{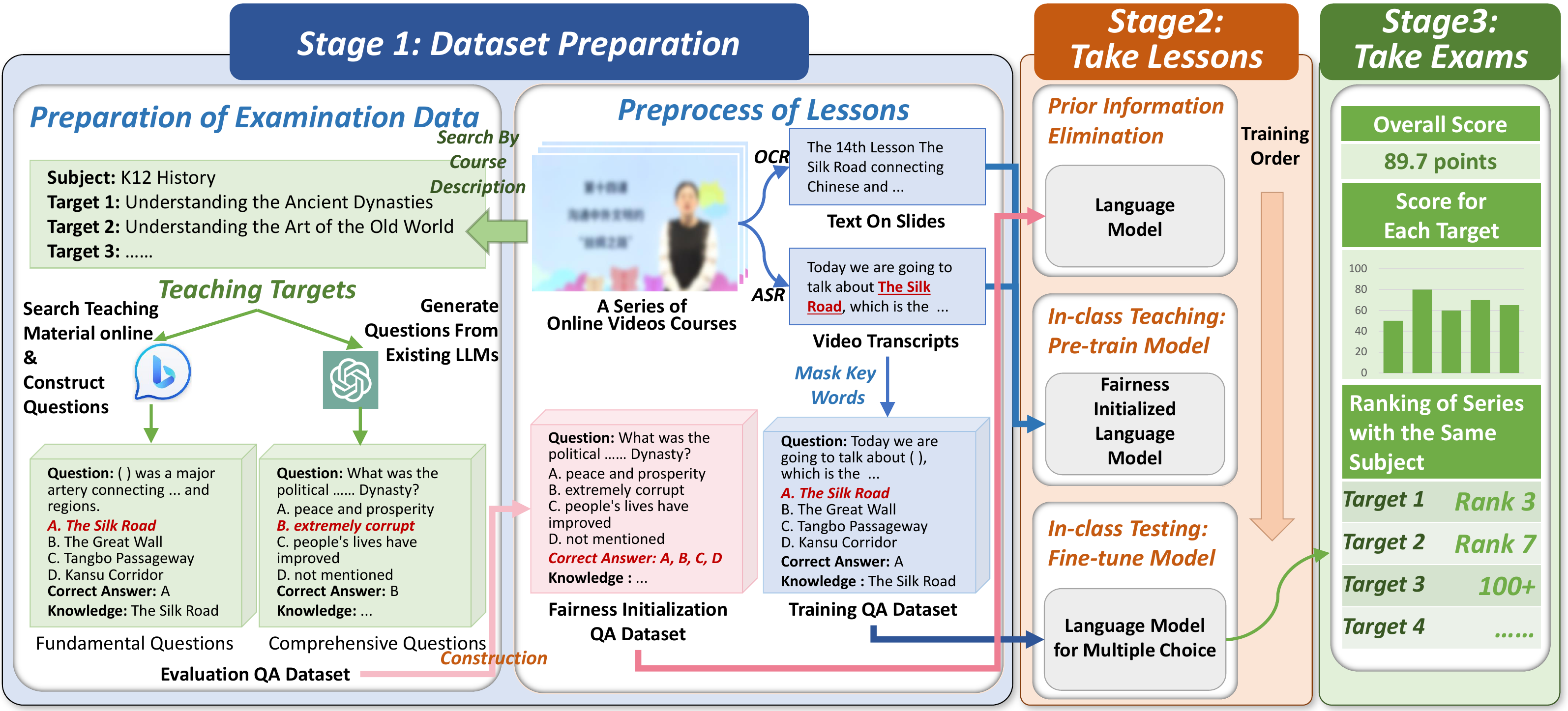}}
\caption{Framework of our proposed VCEval.
VCEval is composed of three main components, which are all detailed in Sec.~\ref{sec:method}: Dataset Preparation in Sec.~\ref{sec:dp}, Take Lessons in Sec.~\ref{sec:tl}, Take Exams in Sec.~\ref{sec:te}.
}
\label{fig:framework}
\end{figure*}

In this section, we elaborate on how the VCEval framework, as depicted in Figure~\ref{fig:framework}, is utilized for the evaluation of video course content.
We focus on elaborating on how the VCEval framework ensures precision, interpretability, and fairness in evaluation.
Given that courses typically contain a substantial amount of textual information, which is the most popular and effective way of conveying knowledge, the primary approach of VCEval is to map the multi-modal information in video courses to text for evaluation and design text-based examination questions for evaluation. The implementation details for each section are presented in the Appendix.

\subsection{Dataset Preparation} 
\label{sec:dp}
Two sets of data should be prepared for VCEval: examination data and teaching data.

\subsubsection{Preparation of Examination Data}
The preparation of examination data entails obtaining the teaching targets $T$ of the video course and the related teaching materials $R(t)$.
We first obtain the teaching targets $T$ by using existing LLMs or search engines based on the course's description, section, and category information.
Then, we use the same method to obtain teaching materials $R(t)$ for each target $t\in T$.

Based on the collected teaching materials, VCEval identifies and masks keywords $y$ within these materials to formulate multiple-choice questions $x$. Additionally, we employ a similarity-based method to sample distractors $O=\{o_1,\dots o_n\}$ for each question $x$ and keyword $y$ from the extracted keywords other than the answer.
Moreover, VCEval utilizes existing LLMs to generate more challenging multiple-choice questions.
Finally, all these generated multiple-choice questions are combined as an evaluation set, which is used to evaluate the model's performance by learning from a variety of video courses:
\begin{equation}
\vspace{-1mm}
R(t)\approx D = \{(x_1,y_1,O_1),\dots,(x_n,y_n,O_n)\}
\vspace{-1mm}
\end{equation}

\subsubsection{Preprocess of Lessons} 
Video courses $V$ mainly contain two different types of information: images $im$ and audio $ad$.
The audio information primarily includes the teacher's explanation of knowledge units.
Since the maturity of existing ASR technology, using existing ASR methods can ensure a high-quality transformation from audio to text.
The image information generally contains a large number of slides, presenting highly structured explanations to knowledge units.
A typical video has at least 24 images per second, so it is necessary to extract representative key images ($im_{key} \subseteq im$) from the video.
VCEval uses the image at the moment when the slide switches, which often contains the richest information among its nearby images, and utilizes OCR technology to obtain the highly condensed text information.
In this way, the preprocess of lessons can be formalized as follows:
\begin{equation}
\vspace{-1mm}
    v \approx OCR(im_{key}) + ASR(ad).
\vspace{-1mm}
\end{equation}
After mapping the multi-model information to text, VCEval also masks the keywords within the video text to construct multiple-choice questions for further fine-tuning the language model.

\subsection{Model Training: Take Lessons}
\label{sec:tl}
During the training phase, our primary objective is to enable the model $\pi_{\theta}(y|x)$ to comprehend and master the knowledge contained in the video course $v$.
The understanding of the video course will bring in an effective evaluation in the subsequent examination phase.
This process can be divided into three main steps: Prior Information Elimination, Learning from In-class Teaching, and Learning from In-class Testing.

\subsubsection{Prior Information Elimination} 
To ensure a fair evaluation, it is necessary to eliminate the influence of the model's random initialization on the accuracy of the testing.
In this step, by taking the advice from model unlearning, VCEval uses examination questions to unlearn the model, which makes the model have the same predictive probability for all candidates in the test.
The formulation is as follows:
\begin{equation}
\vspace{-1mm}
    \arg\min_{\theta}\sum^{\#D}_{j=1}\sum^{\# O_j}_{i=1}P^2(o_{i,j}|x_j;\pi_{\theta}), (x_j, O_j)\in D
    \label{eq:unlearning}
\vspace{-1mm}
\end{equation}
which $\theta$ represents the model parameters, $\#O_j$ denote the option number of examination question $x_j$, and $\#D$ represent the number of all examination questions.
In this way, VCEval ensures that different videos tested using the same benchmark not only have the same initialization state but also have the same predictive accuracy for all options, thereby eliminating the disturbance caused by initialization.

\subsubsection{In-class Teaching: Pre-train Model}
By asking the language model to ``attend lessons'', VCEval uses the text information converted from the video course to pre-train the language model. 
The pre-training process is mainly about predicting the whole lesson from partial information:
\begin{equation}
\vspace{-1mm}
    \arg\max_{\theta}P(v'|v-v';\pi_{\theta}).
\vspace{-1mm}
\end{equation}

\subsubsection{In-class Testing: Fine-tune Model} 
To align with the subsequent examination and to simulate the in-class tests that students undertake in their daily learning to improve their scores, VCEval uses the multiple-choice questions constructed from the video text to fine-tune the model:
\begin{equation}
\vspace{-0mm}
\begin{aligned} 
    \arg\max_{\theta}\sum^{\#D_{ft}}_{i=1}P(y_i|x_i, O_i&;\pi_{\theta}),\\ (x_i, y_i&, O_i)\in D_{ft}
\end{aligned}
\vspace{-2mm}
\end{equation}
Since the pre-trained-only model is incapable of answering multiple-choice questions, the purpose of this step is to make sure the following evaluation based on multiple-choice questions is effective and makes the model understand the video content better.

\subsection{Model Evaluation: Take Exams}
\label{sec:te}
During the evaluation phase, our primary objective is to assess the model's performance in mastering the content of the video course through examinations.
This process mainly includes the target-level evaluation and video-level evaluation.

The evaluation process primarily involves the model answering multiple-choice questions constructed in the Data Preparation Phase.
All questions are related to specific target~(knowledge units or teaching goals), and based on the accuracy of the answers, VCEval will give target-level evaluation by outputting the performance of the video course on each specific target.
By combining all the target-level performances, the video-level performance of the video course can also be obtained.
\begin{equation}
\vspace{-3mm}
\small
\begin{aligned} 
f_{auto}(v,R(t))\approx f_{\pi_{\theta}}(v,D) =&\\ \sum^{\#D}_{i=1}P(y_i|x_i,O_i;\pi_{\theta}),\ (&x_i,y_i,O_i)\in D\\
\end{aligned}
\vspace{-3mm}
\end{equation}

%% file: sources/05.Dataset.02.tex
Given the lack of datasets for video course evaluation, we outline our method for creating a K12 video course assessment benchmark.
Since K12 courses, with their clear teaching targets, are particularly suitable for establishing as an evaluation benchmark, we collected numerous open-source K12 video courses, converting the image and audio information into text using keyframe recognition, OCR, and ASR technologies. 
We also amassed a significant number of K12 textbooks, lesson plans, and syllabi to cover the preset teaching targets. 
We conducted a thorough manual inspection before proceeding with experiments and open sourcing.

\begin{figure}[t]
\centering
\resizebox{\columnwidth}{!}{
\includegraphics{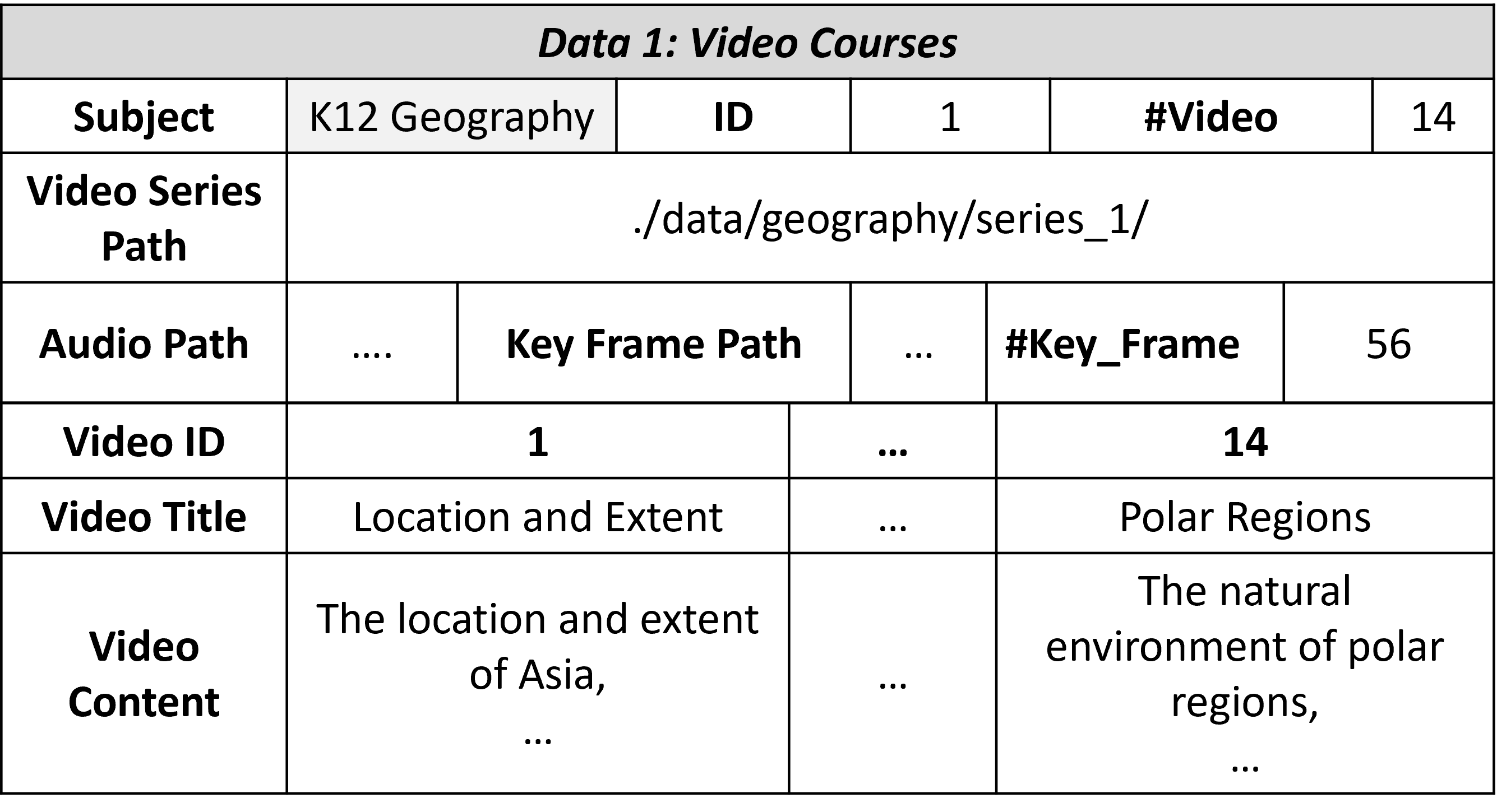}}
\resizebox{\columnwidth}{!}{
\includegraphics{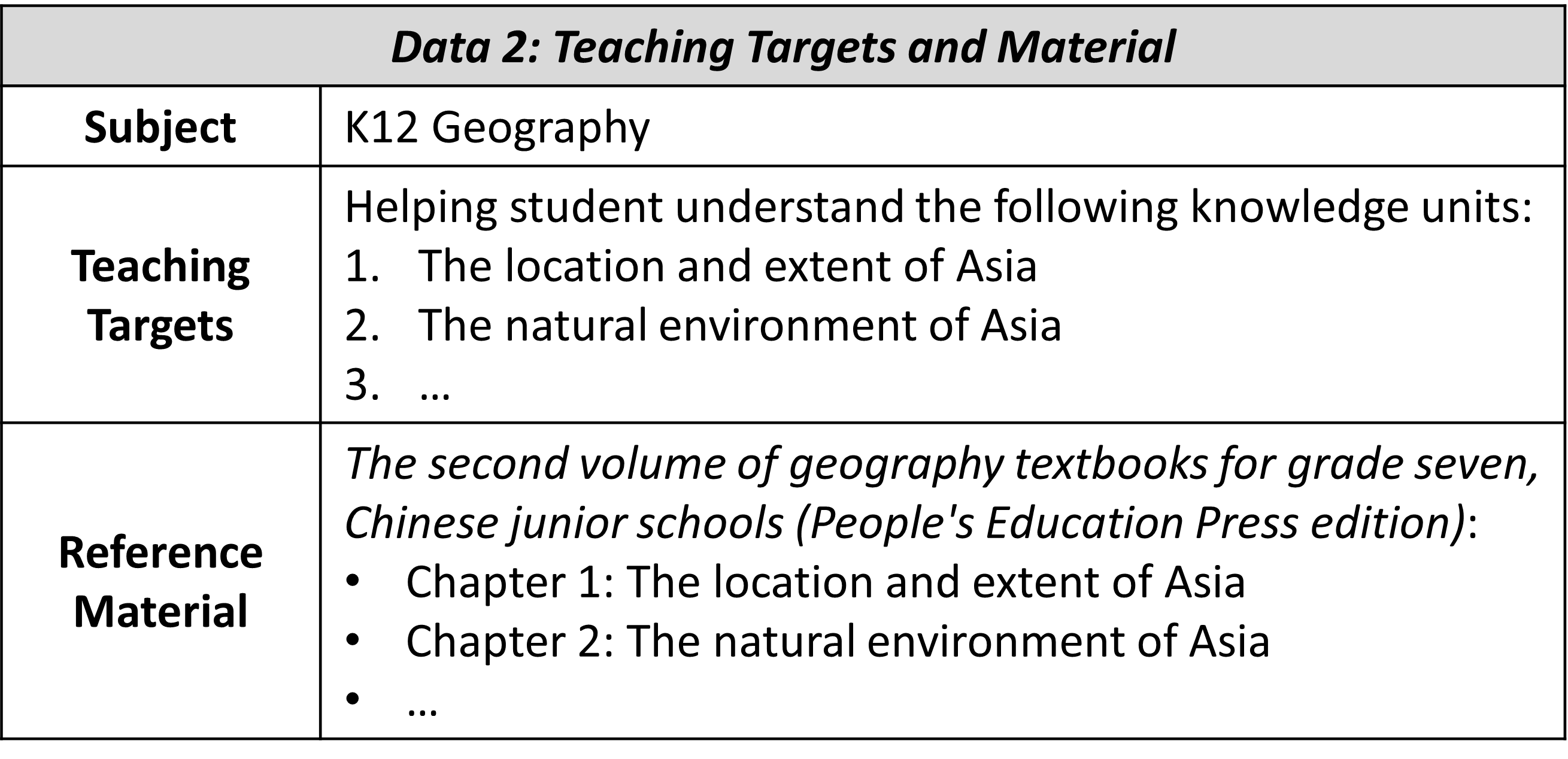}}
\resizebox{\columnwidth}{!}{
\includegraphics{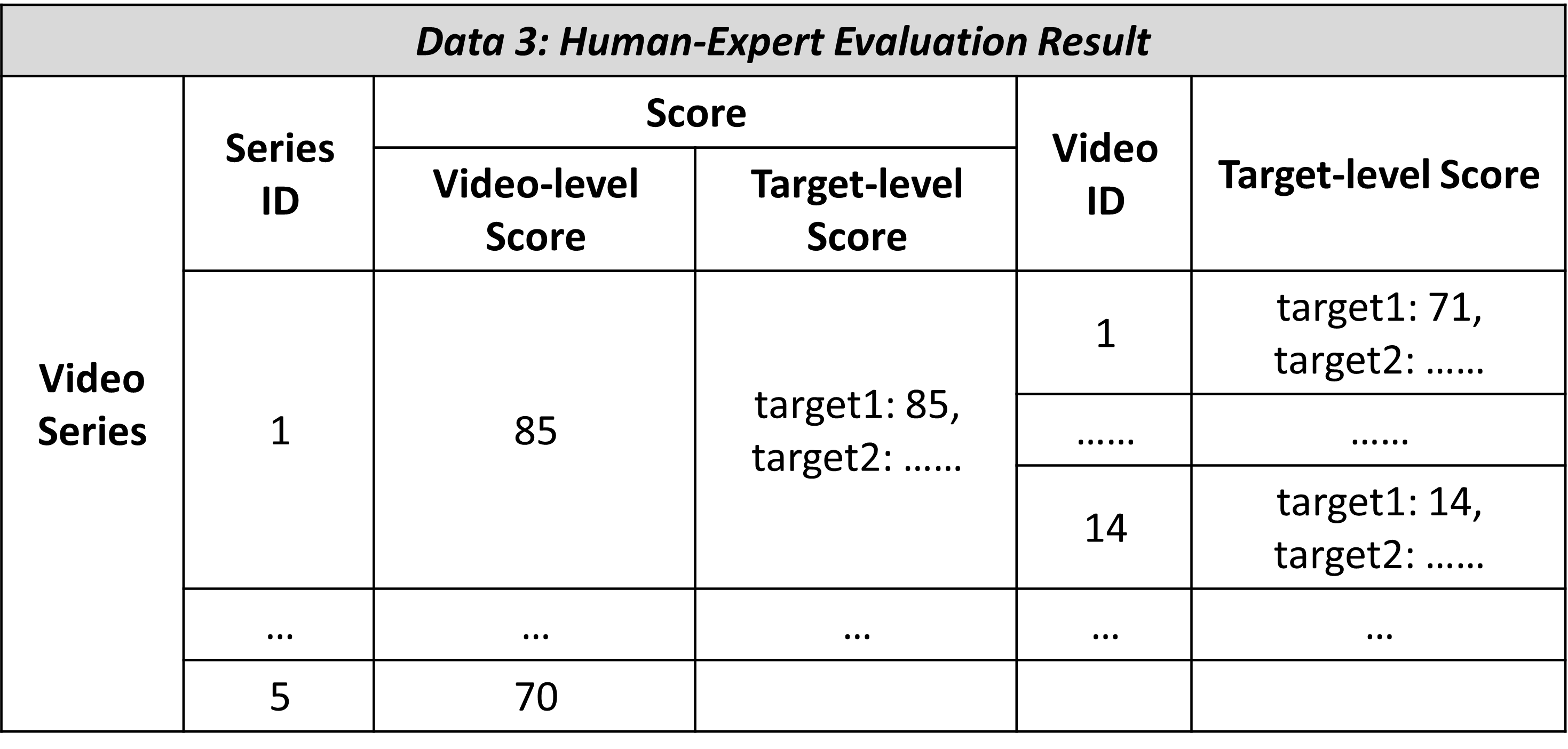}}
\caption{
\textit{Data 1:} An example of a video series. A series contains several videos on a certain subject and each video teaches certain knowledge units. \textit{Data 2:} The collected teaching material with relevant teaching targets. \textit{Data 3:} An example of human annotation.}
\label{fig:series-example}
\end{figure}

\subsection{Video Course Collection}
Video courses often take the form of a series, as shown in the Top of Fig.~\ref{fig:series-example}.
We gathered 15 series of K12 video courses from Bilibili, a leading video-sharing platform in Southeast Asia.
These series focus on three subjects taught in Chinese junior high schools: history, biology, and geography, with five series dedicated to each subject.

Each series within the same subject shares the same teaching targets, which ask them to cover similar content and the same knowledge units, making it feasible to compare their content quality.
The comprehensive statistics of these K12 instructional video series are presented in Tab.~\ref{tab:dataset}. 
The data provide a valuable foundation for further analysis and comparison of the content quality of these instructional videos.
More details about how we transformed these video courses into text data will be listed in the Appendix.

\begin{table}[htbp]
\centering
\begin{tabular}{lccc}
\toprule
\textbf{Subject} & \textbf{History} & \textbf{Biology} & \textbf{Geography} \\
\midrule
$\#Series$ & 5 & 5 & 5\\
\midrule
$\#Videos$ & 101 & 130 & 139\\
\midrule
$\#Minutes$ & 2,733 & 3,218 & 2,808\\
\midrule
$\#Words$ & 585,689 & 652,700 & 615,114\\
\bottomrule
\end{tabular}
\caption{
Brief statistics of the collected video series for each subject.
$\#Minutes$ is the overall duration (minutes) of all videos. 
$\#Words$ is the overall words recognized in all videos.}
\label{tab:dataset}
\vspace{-3mm}
\end{table}

\subsection{Teaching Material Collection}
Furthermore, we have amassed a selection of superior K12 textual pedagogical resources, such as textbooks, from CBook~\footnote{https://github.com/FudanNLPLAB/CBook-150K}, an open-source Chinese book corpus collection, and materials generated by ChatGPT.
These reference materials are meticulously aligned with the predetermined teaching targets and the knowledge units for each subject.
Given that these high-quality resources typically elucidate relevant knowledge units with clarity and precision, fulfilling the teaching targets impeccably, we utilize them to construct examination questions for evaluation, as detailed in the Appendix. 
A comprehensive illustration of the collected data for the K12 video courses is presented in Figure~\ref{fig:series-example}.

\subsection{Manual Annotation}
For the data we have collected, we have primarily conducted manual annotation in the following areas: (1) The accuracy of keyframe identification in the video courses and whether there are instances of information omission due to inaccurate identification. (2) The accuracy of text recognition by ASR and OCR. (3) The filtration and quality refinement of teaching materials. (4) Manual scoring of video courses based on teaching materials. (5) All the ethical problems in data.

We invite the assistance of three graduate students specializing in video comprehension, K12 education, and natural language processing to participate in the manual annotation.
We ensure that the payment for all annotators exceeds the local minimum wage standard.
Additionally, we have set up extra rewards for annotation tasks that require corrections.
Each piece of data~(each sentence in the textual materials and the specific scores for each sub-knowledge point in the video courses) has been annotated by at least two individuals.
All data have been personally reviewed by the first and second authors of this paper after the annotation process.
Detailed annotation specifics can be found in the Appendix.

%% file: sources/06.Experiments.03.tex
In this section, we conduct extensive experiments to show the effectiveness of \textit{VCEval}.
We also present case studies for interpretability.
The experimental settings and ablation studies will be shown in the Appendix.

\subsection{Baselines}
An intuitive evaluation baseline is to compare the textual similarity between video texts and the teaching materials, so several textual similarity evaluation metrics are used as baselines in our experiments.
ROUGE~\citep{lin-2004-rouge} is the most widely used metric in text generation tasks such as summarization and translation, measuring the recall of reference n-grams in the generated texts. We consider three variants \textbf{ROUGE-1}, \textbf{ROUGE-2}, and \textbf{ROUGE-L}.
Conversely, \textbf{BLEU}~\citep{papineni-etal-2002-bleu} is a weighted geometric mean of n-gram precision scores, which also measures the similarity between two texts.

Furthermore, given the recent emergence of highly capable chat models, we use \textbf{ChatGPT} as a strong baseline to evaluate different video texts directly. 
However, the video texts and teaching materials are often too long to input directly into ChatGPT, which is limited by the input length of the model, while VCEval is not limited by the input length.
Thus, we only compare our method with ChatGPT in Section~\ref{pairwise}, which evaluates the segmented clips of a video.
The prompts for evaluation with ChatGPT will be detailed in the Appendix.

For VCEval, we conduct experiments by using BERT~\cite{devlin-etal-2019-bert}~(\textbf{VCEval-BERT}), GPT-2~\cite{radford2019language}~(\textbf{VCEval-GPT}), and BART~\cite{lewis-etal-2020-bart}~(\textbf{VCEval-BART}) as the backbone language model respectively, which represent three different encoder-decoder transformer~\cite{vaswani2017attention} structure.

\subsection{Video-level Experiment}
\label{overall}
The comprehensive performance of VCEval and the baselines are documented in Table~\ref{tab:overall}, while their correlations with manual annotation are detailed in Table~\ref{tab:correlation}.
Across the experiment results, in both scoring and correlation coefficient, VCEval consistently demonstrated results most closely aligned with human evaluation.
However, the other baselines are unable to provide an appropriate assessment.

\begin{table*}[htbp]
\centering
\begin{tabular}{l|c|c|c|c|c|c}
\toprule
    \multirow{2}{*}{Method} & \multicolumn{3}{c|}{Pearson} & \multicolumn{3}{c}{Spearman}\\
    & History & Biology & Geography & History & Biology & Geography\\
\midrule
\multicolumn{7}{c}{\cellcolor{mygray} \textit{{Baselines}}} \\
ROUGE-1 & -0.279 & 0.025 & -0.344 & -0.3 & 0.1 & 0.3 \\
ROUGE-2 & -0.295 & -0.188 & 0.033 & -0.3 & -0.2 & 0.1 \\
ROUGE-L & -0.897 & -0.392 & 0.448 & -0.9 & -0.5 & 0.5 \\
BLEU & -0.105 & -0.180 & -0.070 & -0.3 & -0.3 & 0.1 \\
\midrule
\multicolumn{7}{c}{\cellcolor{mygray} \textit{{VCEval~(Ours)}}} \\
VCEval-BERT & \textbf{0.760} & 0.113 & 0.900 & \textbf{0.7} & 0.2 & \textbf{0.9}\\
VCEval-GPT & -0.078 & 0.073 & \textbf{0.912} & -0.1 & 0.4 & \textbf{0.9}\\
VCEval-BART & 0.117 & \textbf{0.373} & 0.742 & 0.3 & \textbf{0.6} & \textbf{0.9}\\
\bottomrule
\end{tabular}
\vspace{-3mm}
\caption{Pearson and Spearman (\textbf{\textit{the higher the better for both metrics}}) correlation coefficients for three subjects between automated evaluation results and manual annotations.}
\label{tab:correlation}
\vspace{-5mm}
\end{table*}

\begin{table*}[ht]
\resizebox{\textwidth}{!}{
\begin{tabular}{l|c|c|c|c|c|c|c|c|c|c|c|c|c|c|c|c|c|c|c}
\toprule
    \multirow{2}{*}{Method } & \multicolumn{6}{c|}{History} & \multicolumn{6}{c|}{Biology} & \multicolumn{6}{c|}{Geography}& \multirow{2}{*}{Overall}\\
    & Ser.1 & Ser.2 & Ser.3& Ser.4& Ser.5& Ave. & Ser.1 & Ser.2 & Ser.3& Ser.4& Ser.5& Ave. & Ser.1 & Ser.2 & Ser.3& Ser.4& Ser.5& Ave. &\\
\midrule
\multicolumn{20}{c}{\cellcolor{mygray} \textit{{Ground Truth}}} \\
Manual Annotation & 76.0 & 85.0 & 81.0 & 65.0 & 68.0 &\cellcolor{myblue} 75.0 & 89.0 & 81.0 & 70.0 & 63.0 & 60.0 & \cellcolor{mygreen} 72.6 & 95.0 & 62.0 & 79.0 & 84.0 & 60.0 &\cellcolor{myyellow} 76.0 &\cellcolor{mypurple} 74.5 \\
\midrule
\multicolumn{20}{c}{\cellcolor{mygray} \textit{{Baselines}}} \\
ROUGE-1~(x100) & 38.2 & 34.3 & 31.9 & 35.4 & 34.3 &\cellcolor{myblue} 34.8 & 26.3 & 37.1 & 43.3 & 24.9 & 30.3 & \cellcolor{mygreen} 32.4 & 25.2 & 38.7 & 24.9 & 28.2 & 22.9 &\cellcolor{myyellow} 28.0 &\cellcolor{mypurple} 31.7 \\
ROUGE-2~(x100) & 14.7 & 12.9 & 15.1 & 16.7 & 12.2 &\cellcolor{myblue} 14.3 & 11.4 & 14.0 & 27.3 & 12.6 & 13.7 & \cellcolor{mygreen} 15.8 & 8.2 & 12.2 & 8.6 & 11.2 & 5.9 &\cellcolor{myyellow} 9.2 &\cellcolor{mypurple} 13.1 \\
ROUGE-L~(x100) & 21.2 & 15.3 & 17.6 & 22.6 & 26.3 &\cellcolor{myblue} 20.6 & 12.8 & 19.1 & 36.9 & 15.7 & 24.3 & \cellcolor{mygreen} 21.8 & 14.6 & 16.1 & 12.6 & 17.1 & 8.2 &\cellcolor{myyellow} 13.7 &\cellcolor{mypurple} 18.7 \\
BLEU~(x100) & 4.4 & 3.7 & 5.8 & 6.1 & 2.8 &\cellcolor{myblue} 4.6 & 3.3 & 3.9 & 11.5 & 4.1 & 3.9 & \cellcolor{mygreen} 5.3 & 1.4 & 3.3 & 2.1 & 3.2 & 1.0 &\cellcolor{myyellow} 2.2 &\cellcolor{mypurple} 4.0 \\
\midrule
\multicolumn{20}{c}{\cellcolor{mygray} \textit{{VCEval~(Ours)}}} \\
VCEval-BERT & \textbf{69.7} & \textbf{71.3} & 70.8 & 70.2 & \textbf{69.7} &\cellcolor{myblue} \textbf{70.3} & 86.9 & \textbf{82.2} & 89.5 & 87.9 & 80.8 & \cellcolor{mygreen} 85.5 & \textbf{88.3} & 81.2 & 87.0 & 90.4 & 78.1 &\cellcolor{myyellow} 85.0 &\cellcolor{mypurple} 80.3 \\
VCEval-GPT & 62.0 & 63.4 & 66.0 & 64.3 & 64.7 &\cellcolor{myblue} 64.1 & 85.9 & 75.7 & \textbf{84.8} & \textbf{84.6} & \textbf{78.9} & \cellcolor{mygreen} \textbf{82.0} & 85.3 & \textbf{77.0} & \textbf{81.3} & \textbf{86.6} & \textbf{72.1} &\cellcolor{myyellow} \textbf{80.5} &\cellcolor{mypurple} \textbf{75.5} \\
VCEval-BART & 68.9 & 65.9 & \textbf{70.9} & \textbf{65.5} & 69.9 &\cellcolor{myblue} 68.2 & \textbf{87.3} & 83.0 & 86.9 & 87.1 & 80.0 & \cellcolor{mygreen} 84.8 & 86.2 & 82.5 & 83.0 & 88.9 & 82.3 &\cellcolor{myyellow} 84.6 &\cellcolor{mypurple} 79.2 \\

\bottomrule
\end{tabular}
}
\vspace{-3mm}
\caption{The video-level performance of methods in various settings. For VCEval, we report the overall score~(\textbf{\textit{the closer to manual annotation the better}}) on the evaluation dataset for each video course series of each subject.}
\label{tab:overall}
\vspace{-3mm}
\end{table*}

\subsection{Target-level Experiment}
\label{pairwise}
We also conduct an assessment of the video courses in terms of each specific teaching target.
Given that video courses typically provide concentrated teaching towards a particular target, we devised pairwise evaluations for distinct video segments, which is also friendly to manual annotation.
This approach was adopted to ascertain the precision of various models and indicators in evaluating rankings.
There are 53 knowledge units in total for three subjects and we have annotated 530 pairs of data to conduct a pairwise experiment.
We have compared the target-level performance of these video series pairs with different methods and reported the consistency between automated evaluation results and manual annotations in Table~\ref{tab:pairwise}.
For target-level evaluation results, although ChatGPT managed to achieve a respectable score, there remains an average difference of seven percentage points when compared with VCEval.

\begin{table}[t]
\centering
\begin{tabular}{l|c|c|c}
\toprule
Method  & History  & Biology & Geography \\
\midrule
\multicolumn{4}{c}{\cellcolor{mygray} \textit{{Baselines}}} \\
ROUGE-1 & 42.63 & 40.95 & 45.38 \\
ROUGE-2 & 50.00 & 36.19 & 56.15 \\
ROUGE-L & 50.53 & 41.90 & 33.85 \\
BLEU & 50.00 & 39.05 & 50.77 \\
ChatGPT & 65.26 & 67.14 & 65.38 \\
\midrule
\multicolumn{4}{c}{\cellcolor{mygray} \textit{{VCEval~(Ours)}}} \\
VCEval-BERT & 68.42 & 69.52 & 76.92 \\
VCEval-GPT & 60.00 & 69.52 & \textbf{82.31} \\
VCEval-BART & \textbf{72.11} & \textbf{76.19} & 77.69 \\
\bottomrule
\end{tabular}
\vspace{-2mm}
\caption{Pairwise accuracy (\textbf{\textit{the higher the better}}) of methods in various settings. }
\label{tab:pairwise}
\vspace{-3mm}
\end{table}

\subsection{Case Study}
\label{case-study}
\begin{figure}[t]
\centering
\resizebox{\columnwidth}{!}{
\includegraphics{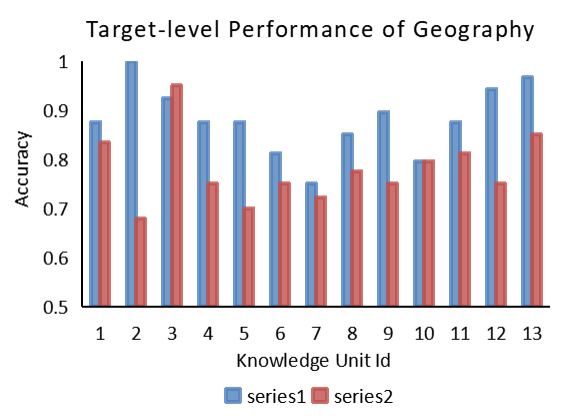}
}
\vspace{-3mm}
\caption{Test accuracy for each knowledge unit of two geography series.}
\vspace{-3mm}
\label{fig:case-bar}
\end{figure}

\begin{figure}[t]
\centering
\resizebox{\columnwidth}{!}{
\includegraphics{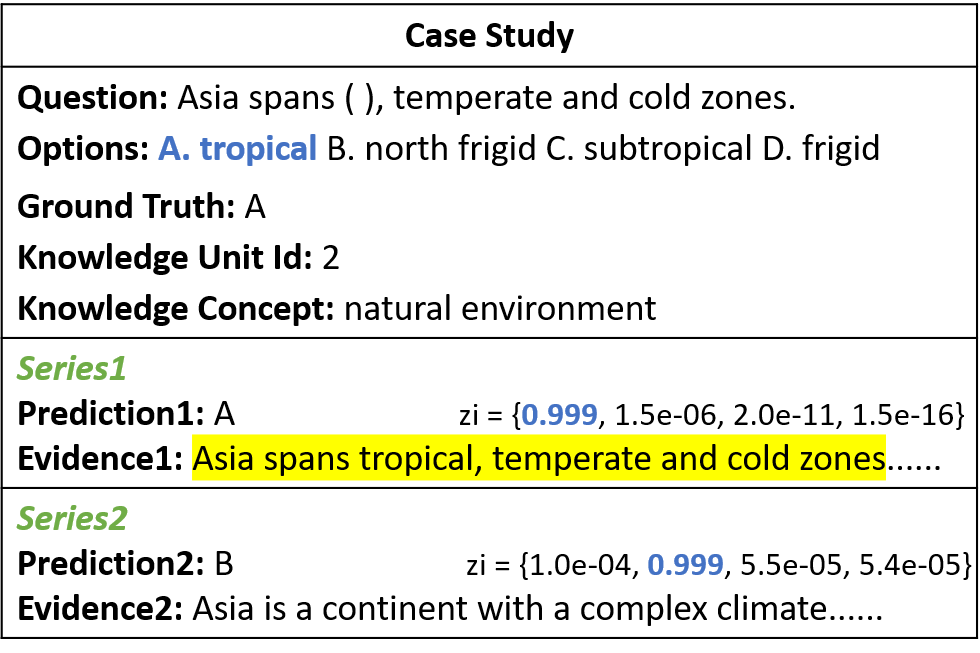}}
\caption{A case study for the result in Figure~\ref{fig:case-bar} about Knowlede Unit Id 2, where the probabilities in prediction $z_{i}$ are probability of option A, B, C, D respectively. The evidences are collected from the video transcripts.}
\label{fig:case}
\end{figure}

We demonstrate the interpretability of VCEval through a case study.
As the case we give in Fig.~\ref{fig:case-bar} and Fig.~\ref{fig:case}, ``Series1'' significantly outperforms ``Series2'' in ``Knowledge2''.
Evidences for this can be found in the transcript of ``Series1'', which contains the phrase ``Asia spans tropical, temperate and cold zones'', whereas ``Series2'' merely mentions ``Asia is a continent with a complex climate''. Consequently, the QA model trained on the text of ``Series1'' correctly answered the question, while the model trained on the text of ``Series2'' failed.

%% file: sources/07.Analysis.02.tex
\subsection{Baselines Shortage}
Using the text similarity~(ROUGE, BLEU) between the video transcript and teaching reference as an evaluation metric showed great disadvantages in our experiments.
We attribute this to: 
(1) the video transcription text contains a large amount of verbal expression, which severely interferes with the calculation of text similarity;
(2) text similarity metrics lack any generalizability in evaluating different expressions about the same meanings; 

The performance of large models~(ChatGPT) is better than text similarity, but still falls short due to the following issues:
(1) the inability to incorporate a large amount of video content, leading to a misunderstanding of the video courses; 
(2) large models pre-train a lot of additional knowledge, which can bias the final evaluation results; 
(3) large models are not adept at numerical scoring evaluations, and their understanding of scores exhibits significant randomness. 
Of course, we believe the biggest drawback of using large models is still the huge overhead brought by pre-training from scratch, as well as the data leakage brought by using existing pre-trained large models.

\subsection{Difference Between Backbone Models}
BERT, GPT, and BART each have their strengths, indicating that the performance of VCEval is backbone model independent.
As to the overall performance, the evaluation results of GPT and BART as backbone models are slightly closer to the manual results than BERT.
This may be due to the combined superiority of the Decoder Transformer itself and the auto-regressive pre-training method~\cite{hoffmann2022training, gu2023xiezhi}.

\subsection{Conclusion}
In this study, we re-evaluate the criteria for an effective educational video and propose three corresponding principles that encapsulate the perspectives of the user, the video uploader, and the online platform administrator.
We introduce the VCEval framework, designed to automatically assess the quality of online video courses in accordance with these principles.
Furthermore, we propose a benchmark grounded in K12 video courses.
Comparative analysis between VCEval and other baselines measuring on this benchmark underscores the superior performance of our proposed methodology.